\documentclass[twocolumn,showpacs,preprintnumbers,amsmath,amssymb]{revtex4}
\usepackage{tabularx,graphicx}\begin{document}
\newcommand{\beq}{\begin{equation}}
\newcommand{\eeq}{\end{equation}}
\newcommand{\beqn}{\begin{eqnarray}}
\newcommand{\eeqn}{\end{eqnarray}}
\newcommand{\bmath}{\begin{subequations}}
\newcommand{\emath}{\end{subequations}}
\newcommand{\bk}{\bold{k}}
\newcommand{\bkp}{\bold{k'}}
\newcommand{\bq}{\bold{q}}
\newcommand{\bkb}{\bold{\bar{k}}}
\newcommand{\br}{\bold{r}}
\newcommand{\brp}{\bold{r'}}
\newcommand{\vp}{\varphi}

\title{Why holes are not like electrons. IV. Hole undressing and spin current in the superconducting state}
\author{J. E. Hirsch }
\address{Department of Physics, University of California, San Diego\\
La Jolla, CA 92093-0319}
 
\date{\today} 
\begin{abstract} 

In paper III of this series we proposed a scenario of superconductivity driven by hole ``undressing'' that involved a complete
redistribution of the occupation of single particle energy levels: the holes near the top of the band were proposed to all condense
to the bottom of the band. Here we consider a  less drastic redistribution involving electrons with a definite spin chirality and show that it is in fact  energetically favored by the Coulomb exchange matrix element $J$ over the scenario proposed earlier. It is shown that spin splitting with chiral states reduces the Coulomb
repulsion and hence that  the Coulomb repulsion promotes spin splitting. Superconductors are proposed to possess a spin-split hole `core' at the bottom of the electronic conduction band
in addition to a spin-split Fermi surface. 
The new 
scenario  leads naturally to the existence of a spin current in the superconducting state
and is consistent with the Spin Meissner effect and negative charge expulsion  discussed earlier within the theory of
hole superconductivity. 

  \end{abstract}
\pacs{}
\maketitle

\section{Introduction}

The theory of hole superconductivity proposes that superconductivity can only occur when the Fermi level is near the top of an electronic energy band
and that it is driven by ``undressing'' of  charge carriers that are hole-like in the normal state and  become electron-like in the superconducting state\cite{sns04}. 
The microscopic interaction proposed to drive this physics is a correlated hopping term $\Delta t$\cite{hole1}  arising from an off-diagonal element of the
Coulomb interaction in a tight binding representation\cite{bondch} in the presence of orbital relaxation\cite{chemp}, that is attractive for states near the top of  a  band and particularly strong when the ions are
negatively charged\cite{diat1}, and that makes the band increasingly more narrow\cite{strong} and incoherent\cite{holeelec1} as the electronic occupation increases.

Analysis of the consequences of this theory for the electrodynamics of superconductors led to the conclusion that drastic modification of the
conventional London electrodynamics is required: a macroscopically inhomogeneous charge distribution is proposed to exist in superconductors, 
with an internal electric field pointing towards the surface\cite{electrodyn}. Associated with it is a spin current flowing within a London penetration depth of the
surface\cite{electrospin}. The process by which the charge inhomogeneity\cite{chargeexp} and the spin current\cite{sm} are generated in the transition from the normal to the superconducting state\
provides a `dynamical' explanation of the Meissner effect\cite{missing}, which does not exist within the conventional London-BCS theory\cite{bcswrong}.

We have also shown in previous work  that the interaction $\Delta t$ yields a lower energy in the presence of a spin current in the superconducting state\cite{sc}.
However, the microscopic origin of the spin current  was not clarified in that work.
Furthermore, in work unrelated to superconductivity we have shown\cite{spinsplit} that another off-diagonal matrix element of the Coulomb interaction, $J$, favors a
new type of Fermi surface instability of metals, 
a  `spin-split state' where  Fermi surfaces of opposite spin electrons shift relative to each other and a spin current develops,
with that state being particularly favored when a band is nearly half-filled. In related work, Wu and Zhang have recently
discussed dynamic generation of spin orbit coupling within a Fermi liquid framework\cite{wuz1} and cast the problem within the general framework of 
Pomeranchuk-type instabilities in the spin channel\cite{wuz2}.

The purpose of this paper is to propose a new scenario for the generation of a spin current within the theory of hole superconductivity.
In the previous paper in this series\cite{holeelec3} we proposed that holes `undress' from the electron-ion interaction by migrating from the top to the
bottom of the band in the transition to superconductivity, and argued that this provides an explanation for the negative charge expulsion and resulting internal electric field 
predicted by the electrodynamics equations. 
However, that scenario did not address the question of how the spin current  develops. Here we show that a   modification of that 
scenario does. Namely, it is not holes of both spin orientations that migrate to the bottom of the band, but only holes of one definite 
chirality. The chirality direction is determined by the Dirac spin-orbit coupling in the presence of the electric field generated by 
charge expulsion, which acts in an analogous way as a `seed' magnetic field would act in a ferromagnet to determine
the preferred broken-symmetry state. A driving force for this effect is shown to be again the off-diagonal matrix
element $J$, which   lowers the energy for a superconductor in this scenario when the band is almost full rather than near half-filling as was the case 
for the  spin-split `normal' state proposed in Ref.\cite{spinsplit}. Furthermore, a key role is proposed to be played by the
ordinary Coulomb interaction and in particular its  long-range nature.

\section{The new scenario}

We start by reproducing a figure from the earlier paper in this series\cite{holeelec3} in Fig. 1, with its caption, depicting
the states of a diatomic molecule with $3$ electrons. We argued in Ref.\cite{holeelec3}
that the Coulomb matrix element $\Delta t$ lowers the energy when two electrons are in an antibonding orbital and hence will
promote electrons from lower to higher orbitals. Then we extrapolated to a lattice system and argued that all the electrons
near the bottom of the band will be promoted to the top of the band, as was depicted in Fig. 6 of ref.\cite{holeelec3}.
This would indeed be the case if the effective hopping amplitude as function of the number of electrons per site $n_e$
\beq
t(n_e)=t_0-n_e\Delta t
\eeq
were to change sign for $n_e$ approaching $2$, the full band. That this is not a totally implausible assumption is shown in Appendix A. However, we believe nevertheless that
it  is too stringent  a requirement.
In fact, for the high $T_c$ cuprates comparison with experiments led us to conclude\cite{hole2} that while $t(n_e)$ becomes very
small as $n_e\rightarrow 2$ it does not change sign.

Nevertheless, for the diatomic molecule the right panel of Fig. 1 also describes schematically the scenario proposed in this paper. 
Which is  that $half$ of the electrons in states near the bottom of the band get promoted to near the top of the band 
when a system becomes superconducting.

  \begin{figure}
\resizebox{8.0cm}{!}{\includegraphics[width=7cm]{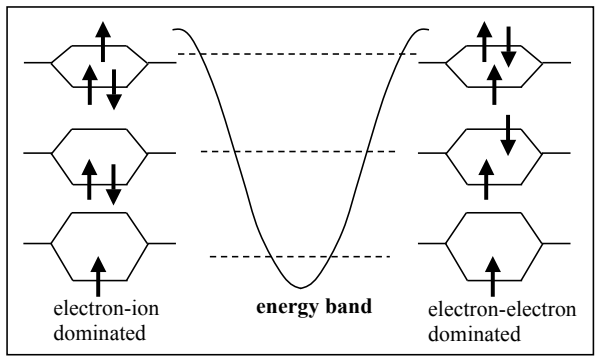}}
  \caption{The diatomic molecule as a microcosm for an energy band. When the electron-electron interaction dominates,
  the occupation of single-particle energy levels changes, as shown on the right side of the figure:
the middle diagram represents a ferromagnet and the upper diagram a superconductor.}
\end{figure}

\section{Spin current and Dirac physics}
Within the theory of hole superconductivity the superconductor expels negative charge from the interior towards the surface\cite{chargeexp}, resulting in 
an excess of negative charge density within a London penetration depth $\lambda_L$ of  the surface\cite{electrospin}
\beq
\rho_-=en_s\frac{\hbar}{4m_e c\lambda_L}
\eeq
with $n_s$ the superfluid carrier density and $\lambda_L$ the London penetration depth given by
\beq
\lambda_L=(\frac{m_e c^2}{4\pi n_s e^2})^{1/2}
\eeq
In the same region near the surface there is a spin current flowing parallel to the surface, with carrier speed given by\cite{sm}
\beq
v_\sigma^0=\frac{\hbar}{4m_e\lambda_L}
\eeq 
The magnitude of $v_\sigma^0$ can be understood from the fact that it gives rise to angular momentum $\hbar/2$ for carriers 
in orbits of radius $2\lambda_L$\cite{sm,missing}, and orbits of radius $2\lambda_L$ are required to
understand the Meissner effect\cite{sm,slater}. 
The carrier spin $\vec{\sigma}$ points parallel to the surface and perpendicular to its velocity $\bold{v}$  according to the relation\cite{electrospin}
\beq
\vec{\sigma}=\bold{v}\times\bold{\hat{n}}
\eeq
with $\bold{\hat{n}}$ the normal to the surface pointing outwards. 

  \begin{figure}
\resizebox{8.5cm}{!}{\includegraphics[width=7cm]{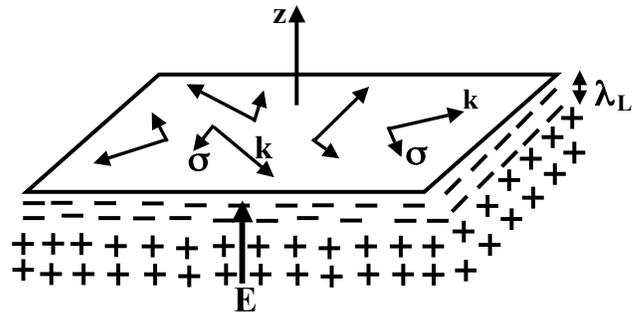}}
  \caption{The plane $z=0$ is the boundary of the superconductor. An electric field in the interior of the
  superconductor points in the $+z$ direction. There is excess negative charge  within distance
  $\lambda_L$ from the surface of the superconductor that was expelled from its interior, and there is a spin current near
  the surface. The long and short perpendicular arrows denote the 
  direction of the momentum and the carrier's spin.}
\end{figure}

Figure 2 shows schematically a superconductor in the region $z<0$ 
bounded by the $z=0$ plane. The Dirac spin-orbit interaction of carriers near the surface in an electric field
$\bold{E}$ pointing towards the surface is\cite{bjorken}
\beq
H_{s.o.}=-\frac{e\hbar}{4m_e^2 c^2}\vec{\sigma}\cdot(\bold{E}\times\bold{p})
\eeq
where $\bold{E}$ is the electric field and $\bold{p}$ is the momentum.  The magnitude of the electric field in the superconductor 
near the surface according to the theory of hole superconductivity\cite{electrospin}  is $E_m$ given by 
\beq
E_m=-\frac{\hbar c}{4e\lambda_L^2}
\eeq
Hence Eq. (6) becomes, with $\bold{p}=\hbar\bold{k}$
\beq
H_{s.o.}=-\frac{\hbar^2}{2m_e} k q_0 (\frac{\hbar}{2m_e c}q_0)\vec{\sigma}\cdot(\hat{k}\times\hat{z})
\eeq
where we have defined
\beq
q_0=\frac{1}{2\lambda_L}.
\eeq
A   spin-orbit term of the form Eq. (8) is commonly referred to as a Rashba term\cite{rashba} and the corresponding spin-split bands as Rashba bands.

The quantity
\beq
r_q=\frac{\hbar}{2m_e c}
\eeq
(proportional to the Compton wavelength)
may be called the `quantum electron radius' (as opposed to the `classical electron radius' $r_c=e^2/m_e c^2$) : a mass $m_e$ orbiting at speed $c$ with radius $r_q$ has angular momentum $\hbar/2$, 
the electron spin, and furthermore the quantum confinement energy of a mass $m_e$ in a
distance $r_q$ is       $\hbar^2/(2m_e  r_q^2)=2m_e c^2$. For a London penetration depth $\lambda_L=160\AA$ ($Al$) the quantity in brackets in Eq. (8) is
\beq
r_qq_0=6.0\times10^{-6}
\eeq
that sets the scale of the spin-orbit symmetry-breaking field arising from Dirac single-particle physics. The spin current speed resulting 
from Eq. (8) is
\beq
v_\sigma^{Dirac}=\frac{\hbar}{2m_e c}q_0 (r_q q_0)=(r_q q_0) v_\sigma^0
\eeq
and the favored spin direction is given by Eq. (5).  

  \begin{figure}
\resizebox{8.5cm}{!}{\includegraphics[width=7cm]{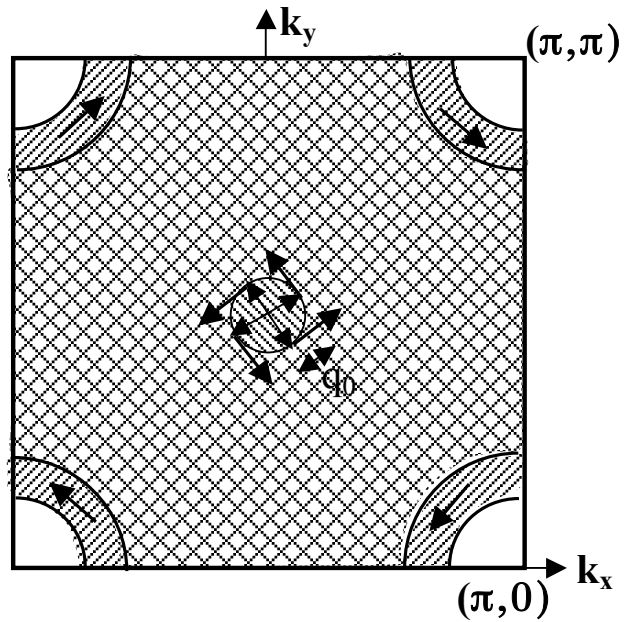}}
  \caption{Schematic image of the Brillouin zone occupation in the superconducting state for an s-like electronic energy band  (bottom of the band at the $\Gamma$-point).
    Depicted is   a $45^\circ$ plane that goes through the 
  points $(0,0,0)$ and $(\pi,\pi,\pi)$ in the Brillouin zone projected onto the $(k_x,k_y)$ plane. The   $\hat{k}_z$ axis points out of the paper.
  A circle of radius $q_0=1/2\lambda_L$ in the $(k_x,k_y)$ plane centered at $(0,0)$ is
  singly occupied by electrons of chirality $\vec{\sigma}\cdot (\hat{k}\times\hat{z})=-1$. The arrows near the center indicate the direction of 
  $(k_x,k_y)$ (radial arrow) and of the spin of the corresponding electron (attached tangential arrow). There is an excess of electrons 
    of chirality $\vec{\sigma}\cdot (\hat{k}\times\hat{z})=1$ at the  Fermi surface in the antibonding regions close to $(k_x,k_y,k_z)=(\pm \pi, \pm \pi, \pm \pi)$, (their
  spin orientation is indicated by the arrows). The resulting regions
  with singly-occupied electrons are denoted by the hatched areas, with $45^\circ$ hatching   (-$45^\circ$ hatching) in the
  antibonding  (bonding) regions respectively.  
  The region hatched by the square pattern is doubly occupied with electrons, and the white regions near $(k_x,k_y,k_z)= (\pm \pi, \pm \pi,\pm \pi)$ are
  doubly occupied with holes.}
  \end{figure}

Thus the situation is similar to that of  of ferromagnetism. The magnetic dipole interaction between electronic
magnetic moments of magnitude $\mu_B$ ($\mu_B$=Bohr magneton) can provide a $qualitative$ explanation of ferromagnetism
(or antiferromagnetism)  but its strength is several orders of magnitude too 
small to explain the
observed ferromagnetism in solids. Heisenberg correctly surmised that the true explanation of ferromagnetism lies in the much stronger Coulomb
interaction between electrons. Similarly here, the spin-orbit interaction arising from the Dirac equation gives rise to a spin current speed
which is $10^{-5}$ times smaller than what the electrodynamic equations of the superconductor tell us it should be\cite{electrospin}, namely Eq. (4).
This is because the ultimate origin of the spin current is again the Coulomb interaction, just as in the case of ferromagnetism.

\section{origin of the spin current and electric field}

In our earlier work on metallic ferromagnetism\cite{metallic} as well as the spin-split state\cite{spinsplit}, we argued that these 
collective effects  are driven
by a Coulomb matrix element $J$ (``bond-charge repulsion''\cite{bc,bc2}) that favors the Fermi surfaces of opposite spin electrons to occupy regions of the
Brillouin zone of {\it opposite bonding character} (i.e. one bonding, one antibonding). We invoke the same physics here.

Consider for definiteness a three-dimensional cubic lattice with an $s-$like band close to full. In real space, the electric field originating in the
charge expulsion points along the $+z$ axis as shown in Fig. 2. According to the Dirac interaction Eq. (8), electrons with spin orientation
satisfying $\vec{\sigma}\cdot (\hat{k}\times\hat{z})=+1$ are slightly favored over those with opposite chirality. We propose that in order to
lower the bond-charge Coulomb repulsion energy, half of the electrons in a small region near $k\sim 0$ (bonding character) will migrate to the
region $(k_x,k_y,k_z)\sim (\pm \pi, \pm \pi,\pm \pi)$. Which half of the electrons? Those with   chirality, $\vec{\sigma}\cdot (\hat{k}\times\hat{z})=1$.
The resulting situation is shown schematically in Figure 3.

As in Ref.\cite{holeelec3}, we argue that the electrons expelled from the center of the Brillouin zone occupying the more costly
electron-ion energy states near the corners of the Brillouin zone   represent the expelled electrons from the interior
of the superconductor that reside near the real space surface. Their chirality corresponds to the direction depicted in Fig. 2 predicted
by the Spin Meissner effect\cite{sm}. 

As Fig. 3 shows, we end up with a new `Fermi surface' deep in the interior of the Brillouin zone, of radius $q_0=1/(2\lambda_L)$. The carriers in that region
of single spin chirality
occupy long-wavelength states ($q_0^{-1}>>$ lattice spacing) that don't `see' the discrete nature of the electron-ion potential, hence they are   `undressed' from the
electron-ion interaction\cite{holeelec2} and behave as free electrons (or holes). They `drive' the behaviour of the entire superfluid condensate in the
superconducting state. The process by which the holes near $(k_x,k_y,k_z)=(\pm \pi, \pm \pi, \pm \pi)$ migrate to $k\sim 0$  \cite{holeelec2}
in the transition from the normal to the superconducting state and acquire chirality is shown schematically in Fig. 4.
It corresponds in 
real space to the orbit expansion from an orbit of atomic dimension (corresponding to a nearly filled
band)  to an orbit of radius $2\lambda_L$, as predicted by the
Spin Meissner effect\cite{sm} and shown schematically in Figure 5. 
 Note that $all$ electrons in the Fermi sea are affected by
this process, which is consistent with the physics of the Meissner effect\cite{meissnerexp}. 
In contrast, within BCS theory only 
a small fraction of electrons near the Fermi surface are affected in the transition to superconductivity.
The orbit expansion is neatly represented by the change in the Larmor diamagnetic susceptibility\cite{missing}
\beq
\chi_{Larmor}(a)=-\frac{n_s e^2}{4m_e c^2} a^2
\eeq
for orbits of radius $a$ from $a=k_F^{-1}$ corresponding to the normal state Landau diamagnetism
\beq
\chi_{Landau}=-\frac{1}{3}\mu_B^2g(\epsilon_F)=\chi_{Larmor}(a=k_F^{-1})
\eeq
(with $g(\epsilon_F)=3n_s/2\epsilon_F$ the density of states)
to the London susceptibility in the superconducting state corresponding to Larmor orbits of radius $a=2\lambda_L=q_0^{-1}$\cite{slater}
\beq
\chi_{London}=-\frac{1}{4\pi} = \chi_{Larmor}(a=q_0^{-1}) .
\eeq

  \begin{figure}
 \resizebox{6.5cm}{!}{\includegraphics[width=7cm]{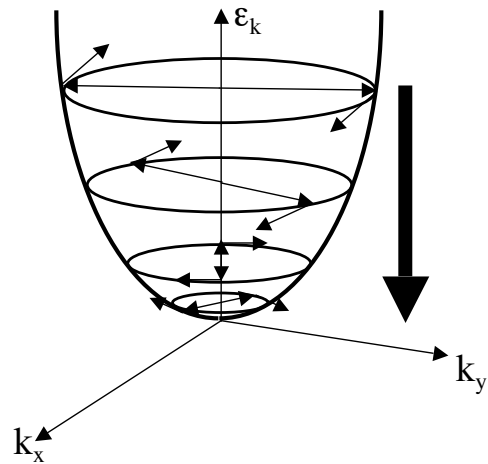}}
 \caption { 
 The figure shows a pair of holes $(\bk\uparrow,-\bk\downarrow)$ with spin orientation $\hat{k}\times\hat{z}$ dropping from the Fermi level to the bottom of the
 electronic energy band. In the process $\bk$ and $-\bk$ rotate around the $\hat{k}_z$ direction and $|\bk|$  shrinks.}
 \label{figure2}
 \end{figure} 

 \begin{figure}
 \resizebox{8.5cm}{!}{\includegraphics[width=7cm]{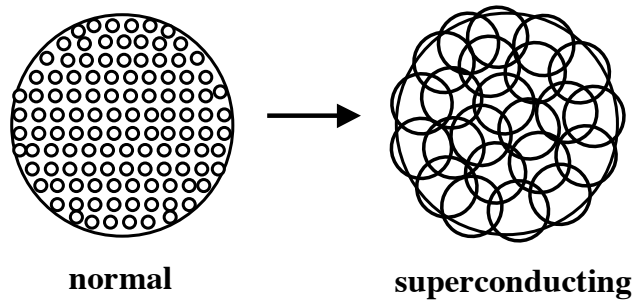}}
 \caption {The figure shows a   superconducting cylinder viewed from the top. 
 Electronic orbits in the normal state have radius $k_F^{-1}$, of order of the ionic lattice spacing, and electronic orbits don't overlap.
 In the transition to superconductivity the orbits expand to radius $2\lambda_L$, several hundreds Angstrom, and they become highly overlapping.
 Only orbits with normal parallel to the cylinder axis are shown. Electrons with spin pointing out of (into) the paper traverse the 
 orbits in clockwise (counterclockwise) direction in the superconductor.
   }
 \label{figure2}
 \end{figure}

The large orbits depicted on the right part of Fig. 5 are highly overlapping, and for that reason phase coherence is a must  to avoid
collisions between electrons that would be very costly in Coulomb energy\cite{emf}. 
This is why a state with the occupations depicted in Fig. 3 cannot exist as a normal metallic state, but requires the long range 
phase coherence characteristic of a BCS-like wavefunction.

\section{Coulomb matrix elements}

Figure 6 shows qualitatively the form of the wavefunction for electrons near the bottom and near the top of a band. 
In the region between the ions the charge density is large for electrons near the bottom of the band (bonding states) and small for electrons
near the top of the band (antibonding states). This is of course the reason for the name: the large charge density between the ions of the bonding
electrons give rise to an effective attraction between the ions, thus bonding the crystal together. When a band becomes almost full with 
many electrons in antibonding states the crystal becomes unstable because of the lack of interstitial charge density in the occupied 
antibonding states.

Figure 6 suggests that the Coulomb repulsion energy will be lowered if some electrons occupy bonding states and some occupy antibonding states,
compared to the case where all electrons occupy bonding state, because the {\it bond charge repulsion}\cite{bc,bc2} between electrons will be
reduced. We have argued that this effect plays a fundamental role in metallic ferromagnetism\cite{metallic} and in the spin-split state\cite{spinsplit},
and we propose here that it also plays a fundamental role in superconductivity, by giving rise to the Brillouin zone occupation
depicted in Fig. 3.

 \begin{figure}
 \resizebox{8.5cm}{!}{\includegraphics[width=7cm]{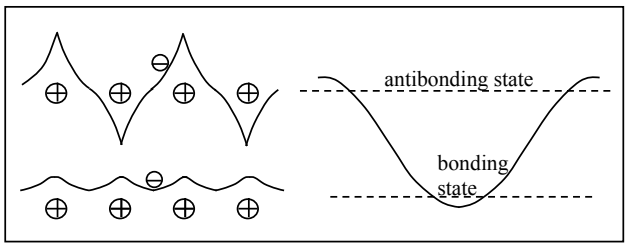}}
 \caption {Qualitative nature of electron energy states near the bottom and near the top of an electronic energy band. 
 For the states near the bottom (bonding states) the charge density is large in the region between the ions, for the states  near the 
 top (antibonding states)  it is small }.
 \label{figure2}
 \end{figure}

 Consider the nearest neighbor   matrix elements of the Coulomb interaction between s-wave like Wannier states $\varphi_i$
\beq
(ij|kl) \equiv \int d^3r d^3 r' \varphi^*_i(r)\varphi_j^*(r')\frac{e^2}{|\bold{r}-\bold{r'}}\varphi_l(r')\varphi_k(r)
\eeq
giving rise to $U=(ii|ii)$, $V=(ij|ij)$, $J=(ij|ji)$, $J'=(ii|jj)$, $\Delta t=(ii|ij)$, all positive. The single particle hopping and the
`hybrid' matrix element $\Delta t$ give rise to the kinetic part of the Hamiltonian\cite{hole1}
\beq
H_{kin}=-\sum_{ij, \sigma} [t_0-\Delta t(n_{i,-\sigma}+n_{j,-\sigma})][c_{i\sigma}^\dagger c_{j\sigma}+h.c.]   .
\eeq
The nearest neighbor exchange interaction $J$ gives for antiparallel spins
\beq
H_J=J\sum_{ij, \sigma} c_{i\sigma}^\dagger c_{j\sigma} c_{j,-\sigma}^\dagger c_{i,-\sigma}
\eeq
and the pair hopping $J'$ 
\beq
H_{J'}=J'\sum_{ij, \sigma} c_{i\sigma}^\dagger c_{i,-\sigma} c_{j,-\sigma}^\dagger c_{j \sigma}
\eeq
Within a mean field decoupling of these terms\cite{metallic,metallic2}, the effective hopping amplitude for an electron of spin $\sigma$ is 
\beq
t=t_0-n_e \Delta t -(J+J') I_{-\sigma}
\eeq
with
\bmath
\beq
n_e=<n_{i\uparrow}>+<n_{i\downarrow}>
\eeq
the site occupation and
\beq
I_{-\sigma}=<c_{i,-\sigma}^\dagger c_{j,-\sigma} >
\eeq
\emath
the bond occupation for the opposite spin orientation.

The matrix elements $J$ and $J'$ involve overlaps of two Wannier orbitals and hence are expected to be negligible beyond
nearest neighbors. The single particle hopping $t_0$ as well as the hybrid matrix element $\Delta t$ involve a single orbital
overlap and thus could be more long-ranged. Hence we assume there is also a next-nearest-neighbor hopping
amplitude
\beq
t_2=t_{02}-n_e\Delta t_2 .
\eeq
As shown in Appendix  A we expect that the ratio $\Delta t_2/t_{02}$ should be larger than $\Delta t/t_0$, and we will
assume that the parameters satisfy $t_0-n_e \Delta t>0$ and $t_{02}-n_e \Delta t_2<0$. For a two-dimensional square   lattice
the dispersion is  then 
\bmath
\beqn
\epsilon_{0k}^\sigma&=&-2t_{nn}(cos(k_x)+cos(k_y)) \nonumber \\
& & +2t_{nnn}cos(k_x)cos(k_y) 
\eeqn
\beq
t_{nn}=t_0-n_e\Delta t -(J+J')I_{-\sigma}
\eeq
\beq
t_{nnn}=n_e \Delta t_2-t_{02}
\eeq
\emath
with $t_{nn}>0$, $t_{nnn}>0$. The bond charge occupation $I_\sigma$ depends on temperature and will increase as the temperature decreases.
We have discussed in earlier work how transitions to ferromagnetic\cite{metallic} and spin-split\cite{spinsplit}
states can occur when $I_\sigma$ reaches a critical value for a half-filled band. Similarly   we envision
here that for a band that is almost full   as the temperature is
lowered $t_{nn}$ decreases relative to $t_{nnn}$ due to the increase in the bond charge occupation $I_\sigma$ and
consequent increase in the electron-electron  bond charge repulsion represented by $J$ and $J'$ and
a point will be reached where it becomes favorable to transfer electrons from $k\sim 0$ to 
$k\sim \pm \pi$. This is shown schematically in Figure 7.

  \begin{figure}
 \resizebox{5.5cm}{!}{\includegraphics[width=7cm]{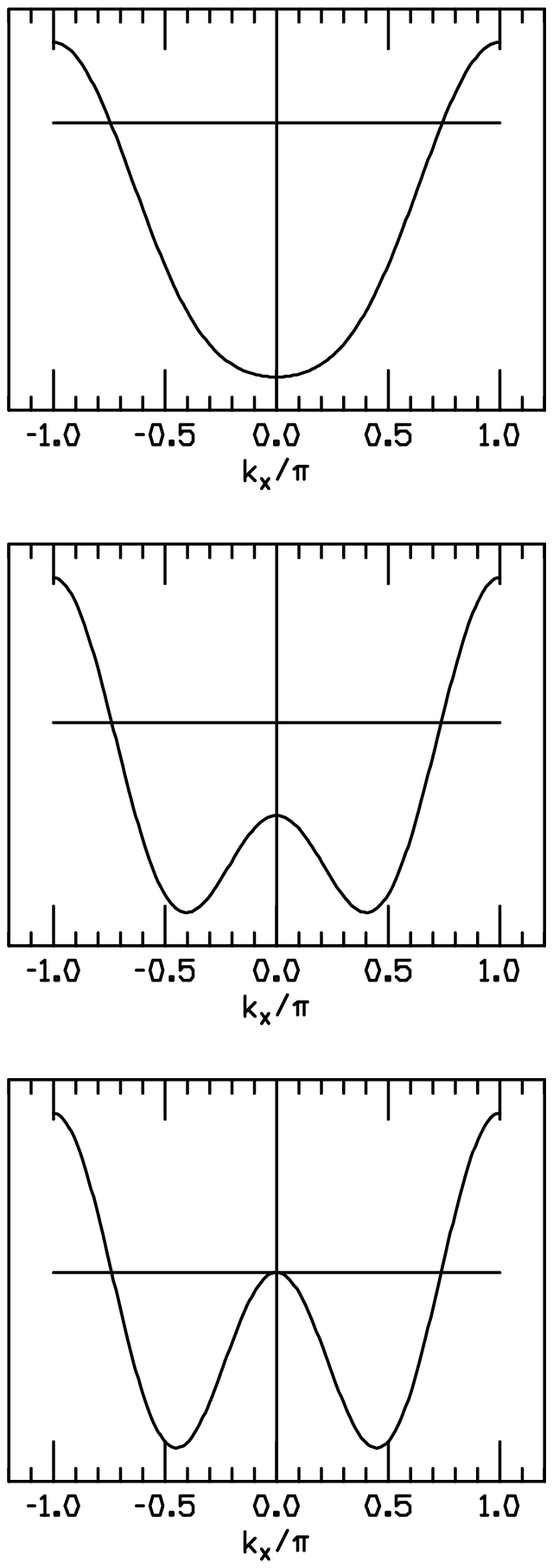}}
 \caption {Evolution of the single particle dispersion (Eq. (23a)  versus temperature
 for a two-dimensional square lattice along the line $k_y=k_x$. The horizontal lines denote the
 position of the chemical potential for band occupation 0.1 holes per site. The parameters for the top, medium and
 bottom panels are       $t_{nn}=1.5$, $0.3$ and $0.16$ with $t_{nnn}=0.5$.
 The bottom panel shows the threshold where holes at the center of the Brillouin zone start to become
 energetically favored.}
 \label{figure2}
 \end{figure}

Below the critical temperature, we expect that the expectation value of the bond charge will break parity, as  in
the spin-split state\cite{spinsplit}. 
For a given $\bold{\hat{k}}$ direction, the bond charge corresponding to
$\vec{\sigma}\cdot (\hat{k}\times\hat{z})=-1$ decreases and the opposite one increases. It can be seen from
Eq. (23b) that when $I_{-\sigma}$ decreases the nearest neighbor hopping amplitude increases and
thus the band energy corresponding to the opposite spin orientation increases.  The spin-split bands will be given by
\beq
\epsilon_{\bk \sigma}=\frac{\hbar^2 k^2}{2m_e} -\frac{\hbar^2}{2m_e}kq_0 \vec{\sigma}\cdot(\hat{k}\times\hat{z})
\eeq
and holes of chirality $\vec{\sigma}\cdot(\hat{k}\times\hat{z})=1$ will condense to the bottom of the lower spin-split band in a circle of radius $q_0$.
A superconducting gap of magnitude\cite{copses} $\Delta_k=\hbar^2 k q_0/2m_e$ opens up at the chemical potential, as shown schematically in Figure 8.

  \begin{figure}
 \resizebox{6.5cm}{!}{\includegraphics[width=7cm]{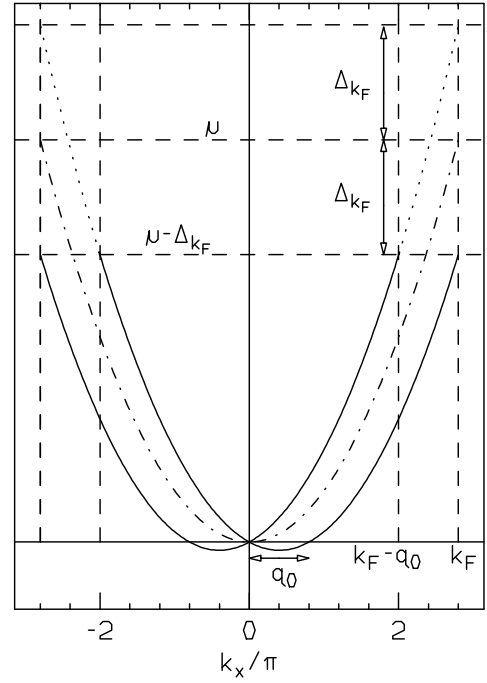}}
 \caption {Spin-split bands below $T_c$ given by Eq. (24). $k_F$ is the Fermi wavevector, and $\mu=\hbar^2 k_F^2/2m_e$. 
 The dot-dashed line denotes the energy dispersion relation in the normal state. The circle of radius $q_0$ in the lower band,
 corresponding to $\vec{\sigma}\cdot(\hat{k}\times\hat{z})=1$,  is
 unoccupied by electrons, i.e. is  occupied by holes.
 Note that in the region $k_F-q_0<k<k_F$ only electrons of chirality $\vec{\sigma}\cdot(\hat{k}\times\hat{z})=1$ exist.}
 \label{figure2}
 \end{figure} 
 
 \section{Coulomb interaction analysis}
 The Coulomb interactions resulting from the Coulomb matrix elements Eq. (16)   are given in momentum space by
 \bmath
 \beq
 H_{UV}=\frac{1}{2N}\sum_{\bk\bkp\bq \sigma \sigma'}   V(\bq) c_{\bk+\bq \sigma}^\dagger c_{\bkp- \bq  \sigma'}^\dagger c_{\bkp   \sigma'} c_{\bk  \sigma}
 \eeq 
  \beq
 H_{J}=\frac{1}{2N}\sum_{\bk\bkp\bq \sigma \sigma'}   J(\bk-\bkp+\bq) c_{\bk+\bq \sigma}^\dagger c_{\bkp- \bq  \sigma'}^\dagger c_{\bkp   \sigma'} c_{\bk  \sigma}
 \eeq 
   \beq
 H_{J'}=\frac{1}{2N}\sum_{\bk\bkp\bq \sigma \sigma'}   J'(\bk+\bkp) c_{\bk+\bq \sigma}^\dagger c_{\bkp- \bq  \sigma'}^\dagger c_{\bkp   \sigma'} c_{\bk  \sigma}
 \eeq 
   \beqn
 H_{\Delta t}&=&\frac{1}{2N}\sum_{\bk\bkp\bq \sigma \sigma'} [\Delta t(\bk) +  \Delta t(\bkp) +\Delta t(\bk+\bq)+ \nonumber \\
& & \Delta t(\bkp-\bq)]  \times   c_{\bk+\bq \sigma}^\dagger c_{\bkp- \bq  \sigma'}^\dagger c_{\bkp   \sigma'} c_{\bk  \sigma}
 \eeqn
  \emath
  with the  interaction amplitudes  given by
  \bmath 
  \beq
  V(\bq)=U+V\sum_\bold{\delta} e^{i\bq \cdot \bold{\delta}}
  \eeq
    \beq
 J(\bq)=J\sum_\bold{\delta} e^{i\bq \cdot \bold{\delta}}
  \eeq
     \beq
 J'(\bq)=J'\sum_\bold{\delta} e^{i\bq \cdot \bold{\delta}}
  \eeq
       \beq
\Delta t (\bq)=\Delta t \sum_\bold{\delta} e^{i\bq \cdot \bold{\delta}}
  \eeq
  \emath
  with $\bold{\delta}$ connecting a site to its nearest neighbors. These interactions are maximally positive for $\bq=0$ and maximally negative for $\bq$ at the
  corners of the Brillouin zone $\bq=(\pm\pi,\pm\pi,\pm\pi)\equiv \pm \vec{\pi}$. Assuming that superconductivity is driven by a reduction of the Coulomb interaction between electrons,
  let us consider the conditions that lead to such reduction for the various terms.

 The self-energy of the electrons is obtained from the  expressions Eq. (25)  for $\bq=0$. From Eq. (25d) it can be seen that $\Delta t$ gives negative contributions
 when $\bk$ and $\bkp$ are close to the corners of the Brillouin zone\cite{mattis,stanford}, thus favoring bands that are almost full and have hole
 conduction in the normal state, in accordance with Chapnik's rule\cite{chapnik}. $J(\bk-\bkp)$ will give negative contributions when
 $\bk-\bkp\sim   \pm \vec{\pi}$, so it favors simultaneous occupation of the center ($\bk\sim 0$) and the
corners    of the Brillouin zone, as in the spin-split scenario discussed in the previous section.
 Similarly $J'(\bk+\bkp)$ will give negative contributions in such a case. $V(\bq=0)$ is uniformly repulsive independent of the occupation of the
 Brillouin zone. 
 
 Next let us considering the scattering terms in the BCS channel:
 \bmath
 \beq
 H_{UV}^{BCS}=\frac{1}{N}\sum_{\bk \bkp}  V(\bk-\bkp) c_{\bk \uparrow}^\dagger c_{-\bk \downarrow}^\dagger  c_{-\bkp \downarrow}  c_{\bkp \uparrow} 
 \eeq
  \beq
 H_{J}^{BCS}=\frac{1}{N}\sum_{\bk \bkp}  J(\bk+\bkp) c_{\bk \uparrow}^\dagger c_{-\bk \downarrow}^\dagger  c_{-\bkp \downarrow}  c_{\bkp \uparrow} 
 \eeq
  \beq
 H_{J'}^{BCS}=\frac{1}{N}\sum_{\bk \bkp}  J'(0) c_{\bk \uparrow}^\dagger c_{-\bk \downarrow}^\dagger  c_{-\bkp \downarrow}  c_{\bkp \uparrow} 
 \eeq
  \beqn
H_{\Delta t}^{BCS}&=&\frac{1}{N}\sum_{\bk \bkp}  [\Delta t(\bk)+\Delta t(-\bk)+\Delta t (\bkp)+ \Delta t(-\bkp)] \nonumber \\
 & &\times
 c_{\bk \uparrow}^\dagger c_{-\bk \downarrow}^\dagger  c_{-\bkp \downarrow}  c_{\bkp \uparrow} 
 \eeqn
 \emath
 The $\Delta t$ terms give negative contribution for scattering when $\bk$ and $\bkp$ are near the corners of the Brillouin zone and thus
 give rise to hole pairing, as previously discussed extensively\cite{hole1,hole2}. 
 $J$ gives negative contributions when $\bk+\bkp$ is close to the corners of the Brillouin zone, hence one pair in the `inner' Fermi surface and the
 other pair in the `outer' Fermi surface in the spin-split state depicted in Fig. 3. $J'$ gives repulsive contribution regardless of the locations of 
 $\bk$ and $\bkp$.
 
 Most significantly, $V(\bk-\bkp)$ will be least repulsive for $\bk-\bkp$ close to $\pm \vec{\pi},$  since the nearest neighbor interaction contribution becomes
 attractive (Eq. (26a)). This fact has been emphasized by Mattis in the context of an electrostatic mechanism for
 superconductivity\cite{mattis}.  In the present scenario,  it will lower the Coulomb energy for pair scattering between the inner and outer Fermi surfaces depicted in Fig. (3).
 
 In summary, we conclude from this analysis that the proposed scenario, with an almost full band in the normal state and spin-splitting as depicted
 in Fig. 3 in the superconducting state, appears to be optimal to reduce the various Coulomb interaction terms arising from 
 the matrix elements of the Coulomb interaction in the Bloch states of the electrons. As discussed elsewhere\cite{molecule,undressing,undressingfm}, the 
 off-diagonal matrix elements of the Coulomb interaction are augmented by orbital relaxation effects which are
 largest for negatively charged ions.
 
 \section{bcs wavefunction and Coulomb interaction}
 We consider a BCS-like wavefunction of the form
 \beq
| \Psi>=\prod_{\bk}' (u_\bk+v_\bk c_{\bk \uparrow}^\dagger c_{-\bk \downarrow}^\dagger)(u_{-\bk}+v_{-\bk} c_{-\bk \uparrow}^\dagger c_{\bk \downarrow}^\dagger)
 \eeq
 where the product is over half the Brillouin zone (e.g. the subspace $k_x>0$) and where we denote by $(\bk\uparrow,-\bk\downarrow)$ pairs of
 favorable chirality ($\vec{\sigma}\cdot(\hat{\bk}\times\hat{z})=1)$ and by
  $(-\bk\uparrow,\bk\downarrow)$ pairs of unfavorable chirality. Quite generally, spin-splitting will be described
 by the condition
 \beq
 |v_\bk| ^2  \neq  |v_{-\bk}|^2   .
 \eeq
 in some region(s) of the Brillouin zone.
 As discussed earlier, this parity breaking would be caused  just by the Dirac spin-orbit interaction in the presence of the electric field generated by
 charge expulsion, but the effect will be greatly augmented by the Coulomb interaction,
increasing  the characteristic wavevector from $q_0\times(r_q q_0)$ to $q_0$.
 
 Consider  scattering processes of the form $(\bk\uparrow,-\bk \downarrow)\rightarrow  (-\bk\uparrow,\bk \downarrow)$ coupling the two Rashba bands. The associated expectation value for
 the Coulomb matrix element $V(\bk-\bkp)=V(2\bk)$ is
 \beq
 V(\bk-\bkp) u_\bk v_\bk^*u_{-\bk}^*v_{-\bk}
 \eeq
 so the interaction is repulsive both for states near the original Fermi surface ($\bk\sim\pm\vec{\pi})$ and near the new small Fermi surface ($k\sim q_0\sim 0$). 
 Spin splitting will reduce this repulsive energy because $u_\bk v_\bk^*$ and $u_{-\bk}^*v_{-\bk}$ will not be simultaneously large for the same $\bk$.
The same is true for other scattering processes coupling both Rashba bands. Thus, in the superconducting state $V(\bk)$ provides a strong driving force for separation of the Fermi surfaces of
opposite chiralities initially split by a tiny amount due to Dirac spin-orbit coupling.

Furthermore for scattering within the same Rashba band $V(\bk-\bkp)$ can give an attractive interaction and corresponding energy lowering. In computing the expectation value
of the Coulomb interaction 
we need to include the overlap matrix elements $M(\bk,\bkp)$ between Rashba spinors, since the orientation of the spin changes with wavevector. For the small Fermi surface the Rashba spinors are
\beq
|\Psi>=   \frac{1}{\sqrt{2}}
   \left(\begin{array}{c}
1\\
-i \frac{(k_x+ik_y)}{\sqrt{k_x^2+k_y^2}}
\end{array}
\right) 
 \eeq
 and the overlap matrix elements for scattering from an initial state $\bk=(q_0,0)$ to $\bkp=(k_x,\pm\sqrt{q_0^2-k_x^2})$ add up to
 \beq
 M_1+M_2=\frac{k_x(k_x+q_0)}{q_0^2}
 \eeq
 so that the interaction is attractive when the angle between $\bk$ and $\bkp$ is between $90^\circ$ and $180^\circ$ ($-q_0\leq k_x \leq 0$).
 Similarly, in scattering processes between the hole pockets near $\pm \vec{\pi}$ involving the same Rashba band, the relative angle between spinors will be 
 (approximately) either $90^\circ$ or $180^\circ$ (see Fig. 3), hence will give vanishing overlap matrix elements
 (corresponding to taking $k_x=0$ and $k_x=-q_0$ in Eq. (32)) and   suppress the repulsion due to $V(\bk-\bkp)$.
 
 For the interaction $\Delta t$ similar effects occur, so in particular the attractive interaction for scattering between hole pockets near $\pm \vec{\pi}$ will be weakened by the
 spin splitting. However, it should be remembered that the magnitude of $\Delta t$ is much smaller than that arising from diagonal elements of the Coulomb interaction, e.g. the 
 nearest neighbor repulsion $V$, so the net effect certainly should be a tendency to enhance spin splitting. For scattering between states near the small Fermi surface
 the interaction $\Delta t$ is strongly repulsive so the beneficial effect of spin splitting also applies to it.
  
 For scattering processes that do not change $(k_x,k_y)$ however this supression of the Coulomb repulsion due to spin splitting does not take place.
 Consider  scattering processes involving the same $\bk-$components in the $xy$ plane and opposite $k_z$ values
 \bmath
 \beq
 \bk=(k_x,k_y,k_z)
 \eeq
  \beq
 \bkb=(k_x,k_y,-k_z)
 \eeq
 \emath
 The expectation value of the Coulomb interaction involving scattering processes $(\bk\uparrow,-\bk\downarrow)\rightarrow(\bkb\uparrow,-\bkb\downarrow)$ 
 (in the same Rashba band) is
  \beq
 V(\bk-\bkb) u_\bk v_\bk^*u_{ \bkb}^*v_{\bkb}= V(2k_z) u_\bk v_\bk^*u_{ \bkb}^*v_{\bkb}
 \eeq
and will be large and repulsive for small $k_z$ since both   $u_\bk v_\bk^*$ and $u_{ \bkb}^*v_{\bkb}$ will be large for $(k_x,k_y)$ on the `small' Fermi surface. To reduce the Coulomb repulsion in this case we suggest that the Bloch wave fermion operators in Eq. (28) should be replaced by the following operators\cite{newbasis}:
\bmath
\beq
\tilde{c}_{\bk\uparrow}^\dagger=\frac{c_{\bk\uparrow}^\dagger+ic_{\bkb\uparrow}^\dagger}{\sqrt{2}}
\eeq
\beq
\tilde{c}_{-\bk\downarrow}^\dagger=\frac{c_{-\bk\downarrow}^\dagger+ic_{-\bkb\downarrow}^\dagger}{\sqrt{2}}
\eeq
\emath
and similarly for the opposite chirality operators. The expectation value corresponding to Eq. (34) with the wavefunction
 \beq
| \tilde{\Psi}>=\prod_{\bk}' (u_\bk+v_\bk \tilde{c}_{\bk \uparrow}^\dagger \tilde{c}_{-\bk \downarrow}^\dagger)(u_{-\bk}+v_{-\bk} \tilde{c}_{-\bk \uparrow}^\dagger \tilde{c}_{\bk \downarrow}^\dagger)
 \eeq
is now zero (except for the particular case $k_z=0$), indicating that this basis will lower the energy\cite{newbasis}. 
 \begin{figure}
 \resizebox{5.5cm}{!}{\includegraphics[width=7cm]{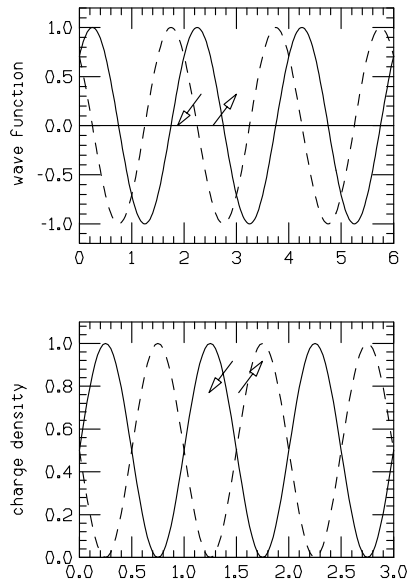}}
 \caption {Wave function amplitudes and charge densities along the $z$ direction for the standing waves associated with the operators  $\tilde{c}_{\bk \sigma}^\dagger$ (full lines) and
 $\tilde{c}_{-\bk \-\sigma}^\dagger$ (dashed lines) , assuming the Bloch states are plane waves. Note that the charge densities avoid each other reducing the Coulomb repulsion.
}
 \label{figure2}
 \end{figure} 
In addition to reducing the Coulomb repulsion for 
scattering processes as seen above, the Coulomb self-energy for pairs $(\tilde{c}_{\bk \uparrow}^\dagger\tilde{c}_{-\bk\downarrow}^\dagger)$
as well as for pairs $(\tilde{c}_{-\bk \uparrow}^\dagger\tilde{c}_{\bk\downarrow}^\dagger)$ is reduced compared to ordinary Cooper pairs,
from $V(0)/\Omega$ to $(V(0)-V(2k_z)/2)/\Omega)$ ($\Omega=$volume). 
This is because these operators create standing waves in the $z$ direction that are $90^\circ$ out of phase between $\bk$ and $-\bk$, as shown schematically in Fig. 9.

The $z$ axis direction corresponds to the direction of the electric field, which is normal to the surface of the superconductor. Thus it is reasonable to use in the BCS wavefunction
Eq. (36) standing waves in the $z$ direction because a superconductor in the ground state or in the Meissner state will carry charge and/or spin current in direction parallel
to the surface but not in direction perpendicular to the surface.  For bodies of more general shape (e.g. spherical or cylindrical) the direction of the normal to the surface
will change with position and the coordinate system needs to be changed accordingly.

Note that the physics discussed here is in some sense opposite to BCS. BCS singled out $(\bk\uparrow,-\bk \downarrow)$ pairs because they provide the largest
phase space for scattering onto other $(\bk'\uparrow,-\bk' \downarrow)$ pairs on the Fermi surface, which is good $provided$ the 
scattering potential is attractive (electron-phonon interaction). Here we have instead a repulsive interaction  and have argued that spin splitting will 
reduce such scattering processes on the Fermi surface for large momentum transfers (eg $(\bk\uparrow,-\bk \downarrow)\rightarrow (-\bk\uparrow,\bk \downarrow)$
and even give negative contribution in a range due to the spin overlap matrix elements.

We point out that it has   recently been shown by Cappelluti, Grimaldi and Marsiglio\cite{frank}  that the tendency to superconductivity  in systems with Rashba
spin-orbit interaction will be greatly enhanced when the Fermi wavevector is comparable to the Rashba wavevector because of an enhanced density of states.  
That is precisely the situation here (Fig. 8), suggesting that  the scattering processes
discussed   involving the `small' Fermi surface will have a dominant effect.

What is the extent of the inner hole pocket in the $k_z$ direction? One can imagine various possibilities for the shape of the small Fermi surface, e.g.
(i) a sphere of radius $q_0$, (ii) a prolate or oblate spheroid with axis along the $\hat{k}_z$ direction,  (iii) a cylinder of radius $q_0$ extending all the way to the edges of the Brillouin zone, or
(iv) a horizontal toroid of inner radius $0$ and outer radius $q_0$\cite{frank}. The answer is not clear at this point. It is also possible that it depends on the
position of the point in real space relative to the surface, which would require a Bogoliubov-De-Gennes description.

The following argument suggests that the inner hole pocket could be a cylinder extending to $k_z=\pm \pi/a$. The cost of expelling electrons from the region
$(k_x,k_y)\sim (0,0)$ to the corners of the Brillouin zone gets progressively smaller as $k_z$ increases as far as the electron-ion and kinetic energy costs are concerned.
The Coulomb energy cost for $V(\bk-\bkp)$ for processes between the inner and outer Fermi surfaces progressively increases. However that cost could be offset
if the gap function $\Delta_\bk$ were to change sign between $\bk\sim(\pi,\pi,\pi)$ and $\bkp=(0,0,k_z')$, rendering a repulsive contribution attractive. In fact, in the model of
hole superconductivity originally considered\cite{hole1,hole2} it was found that $\Delta_k$ changes sign for $\epsilon_\bk$ going down into the band where 
$\Delta t (\bk)$ becomes repulsive, and this provides additional enhancement to $T_c$ (`spatial pseudopotential effect'\cite{hole1,frank2}). In the current scenario we would expect
$\Delta_{\bkp}$ to change sign for large $k_z$ but not for $k_z\rightarrow 0$ since in that case $V(\bk-\bkp)$ is attractive.

Finally, we point out a peculiar feature in our earlier work on the model of hole superconductivity\cite{hole1,hole2}. In a hole representation we had, as in usual BCS
\beq
|v_\bk |^2= \frac{1}{2}    (1-  \frac{\epsilon_\bk-\mu }{E_\bk} )
\eeq
for the hole occupation, with
\beq
E_\bk=\sqrt{(\epsilon_\bk-\mu)^2+\Delta_\bk^2}
\eeq
and the gap function depended linearly on energy
\beq
\Delta_\bk=\Delta_m(-\frac{\epsilon_\bk}{D/2}+c)
\eeq
with $D$ the bandwidth and $\Delta_m$, $c$ constants. For large $\epsilon_\bk$, meaning states deep inside the Fermi sea
\beq
E_\bk\rightarrow\epsilon_\bk   \sqrt{1+(\frac{\Delta_m}{D/2})^2)}
\eeq
and
\beq
|v_\bk|^2 \rightarrow \frac{1}{2}(1-  \frac{1} {\sqrt{1+(\frac{\Delta_m}{D/2})^2)}}) > 0
\eeq
indicating that {\it the hole occupation does not go to zero even for states deep inside the Fermi sea}, unlike conventional BCS theory. An example is shown in Fig. 10, and this
behavior is generic for this model. This could have been seen as a
precursor of the scenario discussed in this paper.
 \begin{figure}
 \resizebox{7.5cm}{!}{\includegraphics[width=7cm]{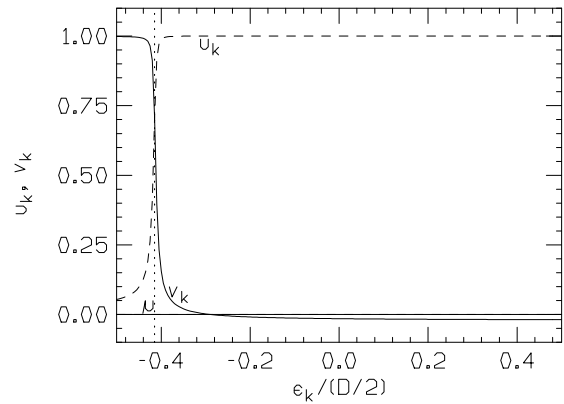}}
 \caption {BCS coherence factors for the correlated hopping Hamiltonian of Ref.\cite{coherence}, in hole representation. The parameters used correspond to Figure 1(a)
 of Ref.\cite{coherence}, with hole occupation $n_h=0.17$ and $T_c=18K$. $D$ is the bandwidth, and the chemical potential $\mu$  is indicated by the dotted line.
 Full and dashed lines are $v_k$ and $u_k$ in hole representation,  the hole occupation is $2|v_k|^2$. 
 Note that the amplitude for hole occupation $v_k$ is non-zero all the way down to the bottom of the electronic band ($\epsilon_k/(D/2)=0.5$).
}
 \label{figure2}
 \end{figure} 

\section{the key to superconductivity}
The arguments in the previous section indicate that: if there is a `small' Fermi surface deep inside an almost full band, it better be spin-split, i.e. corresponding to spins of one 
chirality only. But the arguments don't prove that there is a small Fermi surface. The arguments involving $J$ are suggestive but perhaps unconvincing   for the case of 
nearly-free-electron metals that become superconducting. Furthermore, the arguments given so far do not give indication for why the degree of spin splitting in the
superconductor should be $q_0$.

Consider electron states deep in an almost full  band, with wavevectors of order $q_0$. The Coulomb interaction certainly has a matrix element connecting the state
$(\bk\uparrow,-\bk\downarrow)$ with the state $(-\bk\uparrow,\bk\downarrow)$:
\beq
<-\bk\uparrow,\bk\downarrow|V_c|\bk\uparrow,-\bk\downarrow>=V_c(2k)=\frac{1}{\Omega} \frac{4\pi e^2}{(2k)^2}
\eeq
We have not included in Eq. (42) the  Thomas Fermi wavevector to screen the long-range Coulomb interaction 
as is usually done. We argue that this interaction between states deep in the Fermi sea cannot be Thomas-Fermi screened  
because the electronic states needed to be reorganized in order to screen the interaction are occupied by other electrons and thus not available.
Thus these matrix elements become very large when $k$ becomes very small. 

The reader may argue that because all those states are occupied in the normal state Fermi sea, the Coulomb interaction is unable to scatter $(\bk\uparrow,-\bk\downarrow)$ into
$(-\bk\uparrow,\bk\downarrow)$ because of the Pauli exclusion principle, hence that such processes are irrelevant.

However, there are other states with such wavevectors and spin orientations: the negative energy states in Dirac's theory of the electron.

If the pair $(\bk\uparrow,-\bk\downarrow)$ makes a transition to the top of the band, it opens up room for a pair of electrons in the Dirac sea of negative energy states to make a transition
to that state.

To create a Dirac electron-hole pair costs an enormous energy, $2m_e c^2$. However, the Coulomb matrix element mixing a pair
  $(\bk \uparrow, -\bk \downarrow)$ with another pair $(-\bk \uparrow, \bk \downarrow)$ and $|\bk|=q_0$ has magnitude
\beq
V(2q_0)=\frac{1}{\Omega}\frac{m_e c^2}{n_s}=\frac{m_e c^2}{N_s}
\eeq
with $N_s/2$ the number of superfluid pairs.

Thus, the cost in energy in creating an electron-hole pair from the Dirac sea is comparable to the Coulomb scattering matrix element between pairs of opposite chirality when
there is a single superfluid pair in the system ($N_s=2$).
This suggests that creating Dirac electron-hole pairs plays an important role in the promotion of pairs of one  chirality from the bottom to the top of the
band.  
As more such promotion of half the  pairs  from the bottom to the top of the band occurs, the wavevector  $q_0$ gradually increases and  the Coulomb 
matrix elements decrease  until it becomes too expensive to create Dirac sea pairs and the process stops. The resulting $q_0$ determines the London penetration depth. 

What happens if the  wavefunction   of the superconductor has such an admixture of states with electron-hole pairs created from  the Dirac sea, even if very small?
The new electrons that pop out of the Dirac sea add to the electrons in the metal giving rise to a negative charge density slightly larger than the compensating positive ionic charge density.
What does a metal do with excess negative charge? It pushes it to the surface!\cite{halliday}. This gives rise to the extra negative charge near the surface of
superconductors predicted by the theory of hole superconductivity, shown schematically in Figure 11.
The compensating positive holes created in the Dirac sea instead are uniformly distributed over the volume of the superconductor, giving rise to the interior positive
charge density shown in Figure 11.

 \begin{figure}
 \resizebox{6.5cm}{!}{\includegraphics[width=7cm]{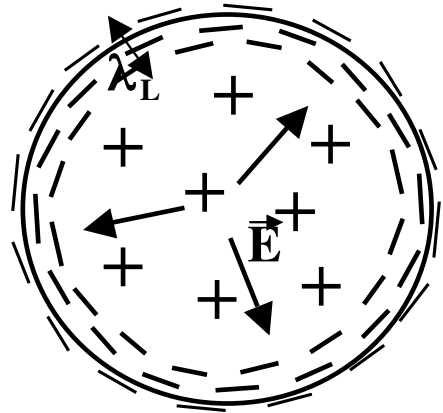}}
 \caption {Schematic picture of a superconducting body (from Ref.\cite{chargeexp}). Negative charge is expelled from the bulk to a surface layer of
 thickness $\lambda_L$, the London penetration depth, and has charge density $\rho_-$ given by Eq. (2).  As a consequence, positive charge density and an outward pointing electric field
 $\vec{E}$  exist in the interior.
 }
 \label{figure2}
 \end{figure} 
 
 For superconductors of linear dimension smaller than the London penetration depth charge expulsion does not occur\cite{chargeexp}, suggesting that electron-hole pairs in the Dirac sea
 are not being  created. 
 This is consistent with the fact that the energy scale of Coulomb repulsion at  distances smaller than $2\lambda_L$ 
 (wavevectors larger than $q_0$) is not large enough to create electron-hole pairs in the Dirac sea. 
For such  a  case the spacing between   $\bk$-points in the Brillouin zone is larger than $q_0$, hence the `small'  Fermi surface in Fig. 3 doesn't really exist, and magnetic fields cannot
be screened.

 It is interesting to note that if the wavefunction of the superconductor has terms with electron-hole pairs created from
 the Dirac sea, the number of electrons is not a good quantum number. The form of the BCS wave function,
 that does not have a definite number of electrons, should have alerted us to this possibility long ago.

If this is correct, the key to superconductivity and the Meissner effect is the long-range nature of the Coulomb repulsion. This was in fact suspected to be the case  in the early days of superconductivity 
research\cite{longrange}. Then, Bohm and Pines\cite{bohmpines} allegedly succeeded in removing the ``troublesome long-range Coulomb interaction"\cite{hoddeson} from consideration in 
metal physics. Perhaps they shouldn't have.

Note that the dispersion relation in the center of the Brillouin zone becomes Dirac-like (Fig. 8), thus mimicking the
behavior of electrons and holes in Dirac theory. The promotion of electrons from $k\sim 0$ to the top of the 
band mimics the promotion of electrons from negative energy states to positive energy states in Dirac theory proposed
here. 

The energy scale of an electronic energy band is a few $eV$, which is of order of  the quantum confinement energy of an electron  in a distance $a_0=\hbar^2/m_e e^2$, the Bohr radius:
\beq
E_{band}=\frac{\hbar^2}{2m_e a_0^2}    .
\eeq
The Dirac energy scale is  the quantum confinement energy of an electron in the quantum electron radius Eq. (10)
\beq
E_{Dirac}=\frac{\hbar^2}{2m_e r_q^2}=2m_e c^2 .
\eeq
The Bohr radius and the quantum electron radius are related by
\beq
r_q=\frac{\alpha}{2}\times a_0
\eeq
with $\alpha=e^2/\hbar c$ the fine structure constant, and the energies are related by
\beq
E_{band}=(\frac{a_0}{r_q})^2E_{Dirac}= (\frac{\alpha}{2})^2 E_{Dirac}
\eeq
The London penetration depth is related to the Bohr radius by
\beq
a_0=\frac{\alpha}{2}\times (2\lambda_L)
\eeq when the superfluid density is $n_s=1/(4\pi a_0^3)$\cite{slater},  
and hence the London penetration depth is related to the quantum electron radius by  
\beq
r_q=\frac{\alpha}{2}\times a_0=\frac{\alpha}{2} \times   \frac{\alpha}{2}\times   (2 \lambda_L)    
\eeq
Thus, the ratio of the band energy scale to the Dirac energy scale is
\beq
E_{band}=(r_q q_0) E_{Dirac}
\eeq
The orbits of electrons in superconductors have radius $2\lambda_L$ and the electrons in the ground state spin current  orbit at speed $v_\sigma^0$ (Eq. 4)\cite{sm}, so
their angular momentum is
\beq
L=m_e v_\sigma^0 (2\lambda_L)=\frac{\hbar}{2}
\eeq
the same as an electron spinning at the speed of light in a circle of radius $r_q$. The kinetic energy
of a spinning electron at speed $c$ can be understood as the quantum zero point energy of a particle confined
to the quantum electron radius $r_q$ and yields its rest mass Eq. (45), similarly the  kinetic energy of an electron in the ground state
spin current in the superconductor can be understood as arising from the quantum confinement of a particle
in a region of linear dimension $2\lambda_L=q_0^{-1}$\cite{sm} 
\beq
\frac{\hbar^2 q_0^2}{4m_e}
\eeq
and we have argued in Ref.\cite{copses} that this   is the condensation energy of the superconductor. 

Thus there is an extraordinary parallel
between electrons in superconductors and real electrons: the superconductor is a giant microscope
dilating length scales by a factor $1/(r_q q_0)=(2/\alpha)^2$,
and the electron in the Cooper pair of the superconductor orbiting at radius $2\lambda_L$ with speed $v_\sigma^0$
is a giant amplified image of the tiny spinning electron itself. 
In a sense (Eq. (49)), the superconductor is a `giant atom'\cite{giantatom} and  the atom is a `giant electron', with `giant$'\equiv2/\alpha$.
It is natural to expect that pair production will play an important role in a giant atom\cite{ga2,ga3}.
The excess negative charge density near the surface $\rho_-$ (Eq. (2)) is the superfluid charge density $en_s$  `diluted' by the expansion factor $(r_q q_0)$ .

\section{Discussion}

In the presence of an electric field, electronic energy bands will spin-split due to the spin-orbit interaction (Rashba splitting\cite{rashba}). Since within the theory of hole superconductivity a macroscopic electric field
exists in the interior of superconductors due to negative charge expulsion, spin splitting of the bands is a necessary consequence. However, we have argued that the spin splitting
that occurs in the superconductor is enormously larger ($\sim 10^5$ times) than that implied by the Dirac spin-orbit coupling with the  macroscopic
electric field and is in fact driven by the Coulomb interaction between
electrons, so it corresponds  to a `dynamic generation of spin-orbit coupling' as proposed by Wu and Zhang\cite{wuz1}.

We have discussed in this paper  a new scenario for superconductivity within the theory of hole superconductivity. This scenario unifies many different elements of the theory, which we
believe makes it compelling. We review these elements below.

(i) The theory predicts that in the transition to superconductivity electronic orbits expand from a microscopic radius to a mesoscopic radius $2\lambda_L$, and that 
a spin current of characteristic wavevector $q_0=1/2\lambda_L$ develops\cite{sm,copses}. In the present scenario this is described by the small spin-split Fermi surface in Fig. 3 
as well as the associated spin-split Fermi surfaces near the corners of the Brillouin zone. The process by which holes migrate from the corners of the Brillouin zone to the bottom of the Brillouin zone in the transition to superconductivity is consistent with the 
real-space picture where the orbits continuously expand\cite{sm}. Note that all electrons in the Fermi sea are affected in both processes.

(ii) In the previous paper in this series\cite{holeelec3} it was argued that occupation of higher $k$-states in the Brillouin zone should be associated with negative charge
expulsion in real space. In the scenario discussed in this paper, this occupation of higher $k$-states near the corners of the Brillouin zone    is directly related to the development
of spin splitting and a spin current, i.e. the expulsion of electrons of one chirality from the center of the Brillouin zone towards the corners. 
In Ref. \cite{electrospin} we derived electrodynamic equations in the charge and spin sectors which showed an intimate relationship between charge expulsion and
spin current development, however the reason for this connection was not intuitively clear from the electrodynamics.

(iii) The Coulomb matrix element $J$ was shown to drive the `normal' spin-split state in ref.\cite{spinsplit}, thus it is natural that it should also play the role in the development
of the spin current that we had proposed exists in all superconductors\cite{sc,sm}. Here we have shown that indeed it plays a key role under the assumptions that spin splitting
occurs at the center of the Brillouin zone and that the normal state Fermi surface is close to the corners of the Brillouin zone.

(iv) We have argued that experiments conclusively show that carriers `undress' from the electron-ion interaction in the transition to superconductivity and behave as
free electrons\cite{holeelec2}. This is consistent with a Fermi surface developing near the center of the Brillouin zone where the states are free-electron-like. 

(v) We have given a series of arguments that the Coulomb repulsion between electrons will be reduced in the scenario discussed here. This is consistent with superconductivity being driven
by the repulsive Coulomb  interaction as proposed in the theory of hole superconductivity rather than by the attractive electron-phonon interaction proposed by BCS.
The Coulomb scattering matrix element $V(\bq)$
between Cooper pairs will quite generally 
be attractive (or least repulsive) for $\bq$ connecting the center and the corners of
the Brillouin zone. 
To take full advantage of this requires Fermi surface parts to reside
near the center of the zone and near the zone corners, as in the scenario discussed here. 
Spin splitting of states with opposite chirality will reduce the Coulomb repulsion for Cooper pair scattering across the Fermi surface and thus the Coulomb repulsion favors spin splitting of the
Fermi surfaces. 
The long range Coulomb repulsion between 
electrons deep in the Fermi sea will be reduced by promoting electrons of one chirality from the bottom to the top of the band, and   electron-hole pairs created from the Dirac sea will take advantage
of it.

(vi) The facts that superconductors require a nearly filled band, that they carry a spin current, and that they expel negative charge from the interior, all being proposed independently
in earlier stages of this theory,  are seen to be naturally related in
the scenario discussed here. Even the fact that holes exist deep inside the electronic Fermi sea was anticipated by the model used in the early stages of the theory\cite{hole1,hole2} (see Eq. (41)), 
even though that prediction of the model was not recognized at the time.

(vii) In the presence of a charge current, the Fermi surfaces in Fig. 3 will be shifted according to the wavevector $\bold{q}$, with $v_s=\hbar q/m_e$ the superfluid velocity.
The phase space for scattering on the small Fermi surface gets progressively smaller and vanishes 
when the edge of the small Fermi surface crosses the point $k=0$, i.e. for $q=q_0$. At that point, which is when the carriers with spin current in direction opposite to the charge current come to
a stop,   superconductivity should be destroyed. This is consistent with the
interpretation of the critical current discussed in connection with the Spin Meissner effect\cite{sm}.

In summary: we have discussed a scenario wherein the topology of the Fermi surface (or  more precisely
its equivalent in the superconducting state)  changes qualitatively in going from the normal to the superconducting state.
We propose that every superconductor has a hidden hole `core', a spin-split Fermi surface enclosing hole-like carriers   in the $k$-space region where the lowest electron states reside in the 
normal state (smooth free-electron-like bonding states\cite{holeelec2}), of radius $q_0=1/2\lambda_L$.
This  core  of undressed hole  carriers with definite chirality and  free-electron-like behavior  `drives' the system and is responsible for  the
remarkable universal physics of superconductors and in particular the Meissner effect as well as the Spin Meissner effect. The detailed form of the BCS wavefunction,
the connection with Dirac physics, and
experimental consequences will be discussed in future work. The proposed hole-core of superconductors may be experimentally
detectable by recently developed high resolution Laser ARPES spectroscopy\cite{laser}.
 
 \appendix
 \section{next-nearest neighbor matrix elements}
 The hopping amplitude between sites $i$ and $j$ for a band arising from local s-orbitals $\varphi_i(\br)$ is
 \beq
 t_{ij}=-\int d^3r  \vp_i^*(\br) [   -\frac{\hbar^2}{2m_e} \nabla^2  + U_i(\br) + U_j(\br)   ]  \vp_j(\br)
   \eeq
   where $U_i(\br)$ is the ionic potential. Assuming $\vp_i$ is an atomic orbital it satisfies $(-\hbar^2/(2m_e)\nabla^2+U_j)\vp_j=\epsilon_j\vp_j$, and furthermore
   assuming $\vp_i$ and $\vp_j$ are orthogonal Eq. (A1) becomes
   \beq
   t_{ij}=-\int d^3r \vp_i^*(\br)U_i(\br)\vp_j(\br)
   \eeq
   We take as ionic potential 
   \beq
   U_i(\br)=-\frac{Ze^2}{|\br-\bold{R_i}|}
   \eeq
   with $Z=2$ the ionic charge, consistent with having a charge-neutral system when the band is full ($n_e=2$). Furthermore we introduce an ionic wave function\cite{coulombatt} $\chi_i(\br)$ which is essentially a $\delta$-function at the position of the ion. The hopping amplitude Eq. (A2) is then
   \beq
   t_{ij}=2\int d^3 d^3r' \vp_i^*(\br)\vp_j(\br)  \frac{e^2}{|\br-\br'|}|\chi_i(\br')|^2
   \eeq
   On the other hand the correlated hopping matrix element $(ii|1/r|ij)$ between these sites is
    \beq
   (\Delta t)_{ij}=\int d^3r d^3r' \vp_i^*(\br)\vp_j(\br) \frac{e^2}{|\br-\br'|}|\vp_i(\br')|^2 
   \eeq
   Comparing Eqs. (A4) and (A5) we conclude that for nearest neighbor matrix elements we will have approximately $2(\Delta t)_{ij}\sim t_{ij}$, i.e. the effective nearest neighbor hopping Eq. (1), 
   $t(n_e)=t_0-n_e\Delta t$ becomes indeed very small as the band becomes full ($n_e\rightarrow 2$). Furthermore, for next-nearest-neighbor sites the fact that 
   $|\vp_i(\br')|^2$ is more spatially extended than $|\chi_i(\br')|^2$ should play a more important role since $\vp_j$ in Eq. (A4) is further away and exponentially decaying,
   hence Eq. (A5) will become relatively larger compared to (A4) for       further than nearest neighbor orbitals. This justifies the assumptions made in 
   Sect. 5 that  the ratio $\Delta t_2/t_{02}$ is  larger than $\Delta t/t_0$ and   that     $t_0-n_e \Delta t>0$ and $t_{02}-n_e \Delta t_2<0$.

 \begin{acknowledgements}
 The author is grateful to Congjun Wu for   helpful discussions.
  \end{acknowledgements}


\begin{references} 

 \bibitem{sns04} J.E. Hirsch, J. Phys. Chem. Solids   {\bf 67}, 21 (2006) and references therein.
   \bibitem{hole1} J.E. Hirsch and F. Marsiglio,  Phys. Rev. B  {\bf 39}, 11515 (1989).
   \bibitem{bondch} J.E. Hirsch, Physica C {\bf 158}, 326 (1989).
   \bibitem{chemp} J.E. Hirsch, Chem.Phys.Lett. {\bf 171}, 161 (1990).
\bibitem{diat1} J.E. Hirsch, Phys.Rev. B{\bf 48}, 3327 (1993).   
\bibitem{strong} J.E. Hirsch, Physica C {\bf 161}, 185 (1989).
\bibitem{holeelec1} J.E. Hirsch, Phys.Rev. B{\bf 65}, 184502 (2002). 
\bibitem{electrodyn} J.E. Hirsch,  Phys.Rev.B {\bf 69}, 214515 (2004).
  \bibitem{electrospin} J.E. Hirsch, Ann. Phys. (Berlin)  {\bf 17}, 380 (2008).
    \bibitem{chargeexp} J.E. Hirsch,  Phys.Rev.B {\bf 68}, 184502 (2003).
  \bibitem{sm} J.E. Hirsch, Europhys. Lett. {\bf 81}, 67003 (2008).
   \bibitem{missing} J.E. Hirsch, J. Phys. Cond. Mat.  {\bf 20}, 235233 (2008).
   \bibitem{slater} J.C. Slater, Phys.Rev. {\bf 52}, 214 (1937).
   \bibitem{bcswrong} J.E. Hirsch, Physica Scripta {\bf 80}, 035702 (2009).
        \bibitem{sc} J.E. Hirsch, Phys.Rev. B{\bf 71}, 184521 (2005).
             \bibitem{spinsplit} J.E. Hirsch,  Phys.Rev.B {\bf 41}, 6828 (1990).
             \bibitem{wuz1} C.J. Wu and S.C. Zhang, Phys.Rev.Lett. {\bf 93}, 036403 (2004).
              \bibitem{wuz2} C.J. Wu, K. Sun. E. Fradkin and S.C. Zhang, Phys.Rev. B {\bf 75}, 115103  (2007).
               \bibitem{holeelec3} J.E. Hirsch,  Int. J. Mod. Phys. B {\bf 23}, 3035 (2009).
               \bibitem{hole2}   F. Marsiglio and J.E. Hirsch,  Phys.Rev. B   {\bf 41}, 6435  (1989).
                 \bibitem{bjorken} J.D. Bjorken and S.D. Drell, {\it Relativistic Quantum Mechanics}, Chpt. 4, McGraw-Hill, New York, 1964.
                  \bibitem{rashba} Y.A. Bychkov and E.I. Rashba, J. Phys. C {\bf 17}, 6039 (1984).
             \bibitem{metallic} J.E. Hirsch, Phys.Rev. B{\bf 40}, 2354 (1989).   
             \bibitem{bc} S. Kivelson, W.P. Su, J.R. Schrieffer and A.J. Heeger, Phys.Rev. Lett.{\bf 58}, 1899 (1987).   
              \bibitem{bc2} D.K. Campbell, J.T. Gammel and E.Y. Loh,  Phys.Rev. B{\bf 38}, 12043 (1988);
              Phys.Rev. B{\bf 42}, 475 (1990).
                      \bibitem{holeelec2} J.E. Hirsch, Phys.Rev. B{\bf 71}, 104522 (2005).   
                      \bibitem{meissnerexp} A.B. Pippard, Phil.Mag. {\bf 41}, 243 (1950).
                      \bibitem{emf} J.E. Hirsch, 
                      	Jour. of Supercond. and Novel Mag., DOI 10.1007/s10948-009-0531-4.
	  \bibitem{metallic2} J.E. Hirsch, Phys.Rev. B{\bf 43}, 705 (1989).   
	   \bibitem{mattis} D.C. Mattis, Int. J. Mod. Phys. {\bf 3}, 389 (1989).
	   \bibitem{stanford} J.E. Hirsch and F. Marsiglio, Physica C {\bf 162-164}, 1451 (1989).
	  \bibitem{chapnik}   I. Kikoin and B. Lasarew, Nature {\bf 129}, 57 (1932); ZhETF {\bf 3}, 44 (1933); Physik.Zeits.d.Sowjetunion {\bf 3}, 351 (1933); A. Papapetrou, Z. f. Phys. {\bf 92}, 513 (1934);
M. Born and K.C. Cheng, Nature {\bf 161}, 1017 (1948); R.P. Feynman {\bf 29}, 205 (1957);
 I.M. Chapnik, Sov,Phys. Doklady {\bf 6}, 988 (1962); Phys.Lett.A {\bf 72}, 255 (1979).
 \bibitem{molecule} J.E. Hirsch, Phys.Rev.B {\bf 48},  3327 (1993).
 \bibitem{undressing} J.E. Hirsch, Phys.Rev.B {\bf 62}, 14487 (2000).
 \bibitem{undressingfm} J.E. Hirsch, Phys.Rev.B {\bf 62},  14131 (2000).
 \bibitem{newbasis} J.E. Hirsch, Phys.Lett. A {\bf 373}, 1880 (2009). Erratum: Phys.Lett. A {\bf 373}, 2800 (2009).
 \bibitem{frank} E. Cappelluti, C. Grimaldi and F. Marsiglio, Phys.Rev.Lett. {\bf 98}, 167002 (2007).
  \bibitem{frank2} F. Marsiglio, Physica C {\bf 160}, 305 (1989).
  \bibitem{coherence} F. Marsiglio and J.E. Hirsch, Phys.Rev.B {\bf 44},  11960 (1991).
 \bibitem{copses} J.E. Hirsch, Journal of Superconductivity and Novel Magnetism {\bf 22}, 131 (2009).
 \bibitem{longrange} W. Heisenberg, Zeits. f. Naturkunde {\bf 2a}, 185 (1947); H. Koppe, Z. Physik {\bf 148}, 135 (1957).
\bibitem{bohmpines}  D. Bohm and D. Pines, Phys.Rev. {\bf 92}, 609 (1953); D. Pines, Phys.Rev. {\bf 92}, 626 (1953).
\bibitem{hoddeson} L. Hoddeson, Journal of Superconductivity and Novel Magnetism {\bf 21}, 319 (2008).
\bibitem{halliday}  D. Halliday, Robert Resnick and J. Walker, ``Fundamentals of Physics'', Wiley, New York, 2001, Chpt. 24.
\bibitem{giantatom} J.E. Hirsch, Phys.Lett. A {\bf 373}, 1880 (2009).
\bibitem{ga2} J.E. Hirsch, J. Phys. Cond. Matt.  {\bf 19}, 125127 (2007).
\bibitem{ga3} B. Muller, Ann. Rev. Nucl. Sci. {\bf 26}, 351 (1976). 
 \bibitem{laser} J.D. Koralek et al, Phys. Rev. Lett. {\bf 96}, 017005 (2006).
    \bibitem{coulombatt} J.E. Hirsch, Phys.Lett. A {\bf 309}, 457 (2003).
               
 \end{references}
 \end{document}